\documentclass[%
 reprint,
superscriptaddress,
 amsmath,amssymb,
 aps,
]{revtex4-2}

\usepackage{graphicx}
\usepackage{dcolumn}
\usepackage{bm}
\usepackage{comment}
\usepackage{graphicx}
\usepackage{dcolumn}
\usepackage{bm}

\usepackage{caption}
\usepackage{amssymb}
\usepackage{amsmath}
\usepackage[dvipsnames]{xcolor}

\usepackage{dcolumn}
\usepackage{bm}
\usepackage{float}
\usepackage{bm}
\def \bF {\pmb{F}}
\def \bH {\pmb{H}}

\def \bd {\pmb{d}}

\def \bd {\pmb{d}}

\def \bx {\pmb{x}}

\def \bY {\pmb{Y}}

\def \bZ {\pmb{Z}}
\def \bw {\pmb{w}}
\def \bW {\pmb{W}}

\def \bv {\pmb{v}}

\def \bdelta {\pmb{\delta}}
\usepackage{rotating}
\usepackage[english]{babel}
\usepackage{amsthm}

\usepackage{subfig}

\makeatletter
\renewcommand{\fnum@figure}{Fig. \thefigure}
\makeatother

\makeatletter
\providecommand{\@LN}[2]{}
\makeatother

\begin{document}


\title{
Synchronization in networked systems with large parameter heterogeneity}

\author{Amirhossein Nazerian}
\affiliation{%
 Department of Mechanical Engineering, University of New Mexico, Albuquerque, NM 87131, USA
}
\author{Shirin Panahi}
\affiliation{%
 Department of Mechanical Engineering, University of New Mexico, Albuquerque, NM 87131, USA
}
\author{Francesco Sorrentino}%
 \email{fsorrent@unm.edu}
\affiliation{%
 Department of Mechanical Engineering, University of New Mexico, Albuquerque, NM 87131, USA
}%

\begin{abstract}
\textbf{Abstract.} 
Systems that synchronize in nature are intrinsically different from one another, with possibly large differences from system to system. 
While a vast part of the literature has investigated the emergence of network synchronization for the case of small parametric mismatches, we consider the general case that parameter mismatches may be large. 
We present a unified stability analysis that predicts why the range of stability of the synchronous solution either increases or decreases with parameter heterogeneity for a given network.
We introduce a parametric approach, based on the definition of a curvature contribution function, which allows us to estimate the effect of mismatches on the stability of the synchronous solution in terms of contributions of pairs of eigenvalues of the Laplacian.
For cases in which synchronization occurs in a bounded interval of a parameter, we study the effects of parameter heterogeneity on both transitions (asynchronous to synchronous and synchronous to asynchronous.)
\end{abstract}

\maketitle
\section{Introduction} \label{sec:intro} 

{Synchronization in networks of coupled dynamical oscillators continues to be the subject of intensive investigation 
\cite{shafiei2019effects,ding2019dispersive,boccaletti2018, sajjadi2020new, awal2020post,fan2021synchronization,palacios2021synchronization}. The emergence of synchronization in homogeneous networks consisting of coupled identical oscillators has been analyzed using the general framework of the master stability function (MSF) \cite{Pe:Ca,boccaletti2014} and the connection graph stability method \cite{belykh2004connection}. 
One of the ongoing challenges is to study the emergence of collective behavior in heterogeneous networks \cite{xiang2007v,ricci2012onset,lu2010cluster,belykh2003persistent,malchow2018robustness,ramakrishnan2022synchronization}, with a large part of the literature on heterogeneous networks focusing on  phase oscillators \cite{KuraBOOK,rosenblum1997phase,lacerda2021heterogeneity,moreno2004synchronization}.}
Typical instances of synchronization in the real world include situations in which the individual parameters of the systems are heterogeneous and are not fine-tuned \cite{wang2012reverse,wiedermann2013node}.
Hence, it is important to study how synchronization may emerge in ensembles of heterogeneous systems with large parameter variability.

Synchronization among non-phase oscillators with parameter mismatches has been studied in a number of papers; however, most of these papers focus on the case of small mismatches.
 In particular, (i) Refs.\ \cite{restr_bubbl,Su:Bo:Ni,SOPO} generalized the master stability approach of Pecora and Carroll \cite{Pe:Ca} to the case of slightly non-identical oscillators. Assuming stability and small parametric mismatches, these papers have found that deviations from the synchronous state grow linearly with the mismatches; (ii) Acharyya and Amriktar \cite{Amritkar2012,Amritkar2015} generalized the approach of (i) by performing a higher-order expansion. The main difference between (i) and (ii) is that (ii) found that the stability is affected by parameter mismatches. Finally,  a recent paper (iii) \cite{sugitani2021} has focused on the case of Chua oscillators and shown that, both in simulation and in experiments, the stability of the synchronous solution is enhanced by mismatches, which contrasts with 
previous observations. Reference \cite{sugitani2021} has suggested that this synchronization enhancement in the presence of mismatches is a general phenomenon that is due to the `mixing' of the modes that describe the time evolution of the perturbations when mismatches are absent. However, it is not clear why such mixing should always result in an improvement for synchronizability. 

In this paper, we provide a unified approach that can be implemented to study the synchronization of a broad class of dynamical networks, for which the stability of the synchronous solution can be written in the form of a linear time-invariant system \cite{tang2022multilayer}. This class includes networks of saddle focus oscillators \cite{Generating2021Ramirez,ROSSLER1976equation}  and networks of piecewise linear oscillators/maps. 
Similar to the classical master stability function approach of Ref.\ \cite{Pe:Ca}, our approach decouples the effects of the dynamics from those of the network topology. However, compared to
 the master stability function approach, we perform a higher order expansion in the heterogeneous parameters and consider the effects of pairs of eigenvalues of the Laplacian (instead of a single eigenvalue) on the stability of the synchronous solution. {Our approach allows us to explain why parameter heterogeneity sometimes hinders and sometimes favors network synchronization,} which was not considered in \cite{sugitani2021}.

\section{Results}

\subsection{Parametric Mismatches in the Local Dynamics} \label{sec:local}
We consider a general equation for the time evolution of $N$ coupled dynamical systems, with parametric mismatches in the individual dynamics of each node,
\begin{equation} \label{main:gen}
\dot{\bx}_i(t)=\bF(\bx_i(t),r_i)-\sigma \sum_j L_{ij}\bH(\bx_j(t)),
\end{equation}
$i=1,\hdots,N$, $\bx_i \in \mathbb{R}^m$ is the dynamical state of the node $i$, $\bF : \mathbb{R}^m \mapsto \mathbb{R}^m$ and $\bH : \mathbb{R}^m \mapsto \mathbb{R}^m$ are the local dynamics of each node and the coupling functions between the nodes, and $\sigma$ is a positive scalar measuring the strength of the coupling. The symmetric Laplacian matrix $L=[L_{ij}]$ is constructed as $L = G - A$, where $A = [A_{ij}]$ is the adjacency matrix that describes the network topology, and $G$ is a diagonal matrix where the diagonal entries are the sums of the rows of $A$. Each parameter is set equal to $r_i=\bar{r} +\epsilon\delta_i$, where $\bar{r}=N^{-1}\sum_{i=1}^N r_i$ is a nominal average value and $\epsilon\delta_i$ is a parametric mismatch, with $\sum_j \delta_j=0$, and $\sum_j \delta_j^2=1$; the scalar $\epsilon$ is the tunable magnitude of the mismatches.

{Our work applies to different classes of the master stability function \cite{Report}, i.e., to Class II, for which stability of the synchronous solution is achieved in an infinite range of the coupling gain $\sigma$ and to Class III,  for which stability of the synchronous solution is achieved in a finite range of the coupling gain $\sigma$. For Class II,
as we increase $\sigma$ from zero, there is only one transition from asynchrony to synchrony (A $\rightarrow$ S). For Class III, as we increase $\sigma$ from zero, first there is  a transition from asynchrony to synchrony (A $\rightarrow$ S) followed by another transition from synchrony to asynchrony (S $\rightarrow$ A.) We develop a theory that allows us to study the effects of parameter mismatches on each individual transition and applies to either the case of A $\rightarrow$ S transitions or S $\rightarrow$ A transitions.} 

We assume possibly large mismatches and small perturbations about the average solution. The particular assumption of small perturbations is discussed in detail in Supplementary Note 1.
By defining $\Bar{\bx}(t)=\frac{1}{N}\sum_i \bx_i(t)$ and $\Bar{r}=\frac{1}{N}\sum_i r_i$, we can write an equation for the average solution and the time evolution of small perturbations ${\bw}_i(t)= {\bx}_i(t)- \bar{\bx}(t)$,
\begin{subequations}
\begin{equation}
    \dot{\bar{\bx}}  = \frac{1}{N} \sum_j \Bigl(\bF(\bar{\bx}, r_j) + D\bF (\bar{\bx}, r_j) \bw_j \Bigr)
\end{equation}
\begin{align}\label{der:gen}
\begin{split}
    \dot{\bw}_i = & D{\bF}(\Bar{\bx},r_i) \bw_i  - \sigma \sum_j L_{ij} D \bH(\Bar{\bx}) \bw_j + \bF(\bar{\bx}, r_i) \\
    & -\frac{1}{N} \sum_j D\bF (\bar{\bx}, r_j) \bw_j -\frac{1}{N} \sum_j \bF(\bar{\bx}, r_j)
\end{split}
\end{align}
\end{subequations}
\color{black}
where $D$ is the differential operator. {We emphasize that the derivations we present only apply to the case that the perturbations in the states of the systems are small. At the same time, we do not introduce an assumption of small mismatches.
}

In the following, {for each transition, either  A $\rightarrow$ S or S $\rightarrow$ A,} we investigate how parameter mismatches affect the network `synchronizability' \cite{Ba:Pe02}, i.e., the range of the parameter $\sigma$ over which the synchronous solution is stable. When we say that mismatches improve (hinder) synchronization, we mean that {for that particular transition} this range is extended (narrowed.) {In particular we study separately, each one of the transitions that may arise, either (A → S) or (S → A) and see how the critical $\sigma$ that characterizes those transitions is affected by parameter heterogeneity. We remark that this analysis needs to be repeated for each individual transition 
and in fact, in what follows we report the example of a particular system that has two transitions, A $\rightarrow$ S and S $\rightarrow$ A,  and for which parameter heterogeneity increases the synchronizability in the case of the A $\rightarrow$ S transition and reduces it in the case of the S $\rightarrow$ A transition.}

We proceed under the assumption that the particular choice of the local dynamics $\bF(\cdot)$ and of the coupling function $\bH(\cdot)$ results in a weak dependence of the Jacobians on the average solution, 
see \cite{tang2022multilayer}. 
This assumption has been previously validated for saddle-focus oscillators (e.g. the R\"{o}ssler system) in \cite{tang2022multilayer} and is validated here for the Chua circuit and Bernoulli maps, see Supplementary Note 2 for a discussion on the validity of this assumption. 
We can then write $D{\bF}(\Bar{\bx}(t),r_i)\approx D{\bF}(r_i)$, where $D{\bF}(r_i)= F +\epsilon\delta_i B$, where $F = D{\bF}(\bar{r})$ and $B \in \mathbb{R}^{m \times m}$ are constant matrices. {Analogously, we can also write $D{\bH}(\Bar{\bx}(t))\approx H$, where  $H \in \mathbb{R}^{m \times m}$ is a constant matrix.}
We comment further on this assumption. While the assumption only applies to certain choices of the functions $\bF$ and $\bH$, we will show successful application of the theory to a variety of continuous time and discrete systems, such as R\"{o}ssler systems, Chua circuits, Bernoulli maps, and opto-electronic maps, as we show in the rest of this paper. More importantly,  
we will be able to address the case of large parameter mismatches and to analyze whether parameter mismatches either enhance or hinder synchronization.

We now define $\bW = [\bw_1^\top \,\, \allowbreak \bw_2^\top \,\, \allowbreak \hdots \, \allowbreak \bw_N^\top]^\top$, and rewrite the homogeneous part of Eq.\,\eqref{der:gen} in compact form,
\begin{align} \label{eq:hom}
\begin{split}
    \dot{\bW}(t) = M \bW(t),
\end{split}
\end{align}
where $M=M_0+\epsilon M_1$, $M_0=I_N \otimes F -  \sigma ({L} \otimes H)$  {with $I_N$ indicating the identity matrix of size $N$}, and $M_1= \Delta \otimes B$, $\Delta = \textnormal{diag}({\delta}_1, ..., {\delta}_N)$.

It is important to emphasize that while Eq.\,\eqref{der:gen}
has both a homogeneous part and a non-homogeneous part, we are mostly concerned with the homogeneous part of the equation. 
The homogeneous part of Eq.\,\eqref{der:gen} describes stability;  in case the homogeneous part is stable, the non-homogeneous part quantifies the deviation from the synchronous state. The rest of this paper is devoted to analyzing how parametric mismatches affect the homogeneous part of the equation. This is in contrast with Refs.\ \cite{restr_bubbl,Su:Bo:Ni,SOPO} that focused on the non-homogeneous part of the equation and did not study how parametric mismatches affect the synchronizability but only the level of synchronization achieved in the network in case the synchronous solution is stable. 
In order to capture the dynamics of the transverse perturbation about the average solution, we introduce the $N \times m$-dimensional matrix $\tilde{V}$, with $m=(N-1)$, that has for columns all the eigenvectors of the matrix $L$ except for the eigenvector $[1,1,...,1]$, $\tilde{V}^\top \tilde{V}=I_m$.
 We have that $\tilde{V}^\top L \tilde{V}$ is equal to the $m$ dimensional matrix $\tilde{\Gamma}$, with entries on the main diagonal, $-\gamma_m \leq -\gamma_{m-1} \leq \cdots \leq -\gamma_1 < 0$.
 By multiplying Eq.\,\eqref{eq:hom} on the left by $(\tilde{V}^\top \otimes I)$ and on the right by $(\tilde{V} \otimes I)$ and obtain $\tilde{M}=\tilde{M}_0+\epsilon \tilde{M}_1$, $\tilde{M}, \tilde{M}_0, \tilde{M}_1 \in \mathbb{R}^{nm \times nm}$, where
\begin{subequations} \label{eq:mtildelocal}
\begin{equation}
    \tilde{M}_0 = (\tilde{V}^\top \otimes \tilde{V}) M_0 (V \otimes I)= I_{m} \otimes F - \sigma \tilde{\Gamma} \otimes H,
\end{equation}
\begin{equation}
    \tilde{M}_1 = (\tilde{V}^\top \otimes I) M_1 (\tilde{V} \otimes I) = \tilde{\Delta} \otimes B,
\end{equation}
\end{subequations}
and we have used the property that $V^\top \Delta_2 V=V^\top {D} V= 0$.
The transformed matrix $\tilde{\Delta}= V^\top \Delta_1 V \in \mathbb{R}^{m \times m}$ is symmetric and has a zero trace.

Next, we analyze stability of the transverse perturbations about the average solution by studying the sign of the eigenvalues of $\tilde{M}$ as a function of the magnitude of the mismatches, $\epsilon$. We use matrix perturbation theory to estimate the variation of the eigenvalues of $\tilde{M}$ as a function of $\epsilon$. This will result in a parametric approach, similar to that of the master stability function \cite{Pe:Ca}, to analyze the effect of mismatches on the synchronizability of  networks of oscillators with mismatches. 

We call $\Lambda$ the matrix that has the eigenvalues of $\tilde{M}$ on its diagonal, and by using second-order matrix perturbation theory \cite{bamieh2020tutorial}, we approximate $\Lambda(\epsilon)=\Lambda_0+\epsilon \Lambda_1 +\epsilon^2 \Lambda_2$, where $\Lambda_0$ is a diagonal matrix with the eigenvalues of $\tilde{M}_0$ on the main diagonal, and $\Lambda_1$ and $\Lambda_2$ are diagonal matrices to be calculated as a function of $\tilde{M}_0$ and $\tilde{M}_1$.

Given the block diagonal structure of $\tilde{M}_0$,
 its left ($W_0$) and right ($V_0$) eigenvectors are written in the form $W_0=\oplus_{i=1}^{m} W_0^i$ and $ V_0=\oplus_{i=1}^{m} V_0^i$ where
${W_0^i}^* V_0^i=I$, and the superscript $*$ indicates the conjugate transpose. 
We define:
\begin{equation*}
\begin{gathered}
    \Lambda = \oplus_{i=1}^{m} \Lambda_i, \quad \Lambda_i = \textnormal{diag}({\lambda}_{i}^1, \hdots, {\lambda}_{i}^m) \in \mathbb{C}^{m \times m} \\
    \Lambda_\dagger = \oplus_{i=1}^{m} \Lambda_{\dagger,i}, \quad \Lambda_{\dagger,i} = \textnormal{diag}(\lambda_{\dagger,i}^1, \hdots, \lambda_{\dagger,i}^m) \in \mathbb{C}^{m \times m}
\end{gathered}
\end{equation*}
where $\Lambda_\dagger$ refers to any of the $\Lambda_0, \Lambda_1$, or $\Lambda_2$ matrices. 
The left ($\bw_{0,i}^j$) and the right ($\bv_{0,i}^j$) eigenvectors corresponding to the eigenvalue $\lambda_{0,i}^j$ are the column $j$ of $W_0^i$ and $V_0^i$, respectively.
We obtain,
\begin{subequations}
\begin{gather}
    {\Lambda_1=\mbox{diag}({W_0^*}\tilde{M}_1{V_0})=\mbox{diag}\left({W_0^*} (\tilde{\Delta} \otimes B) V_0\right)}\\
    \Lambda_2 = - \text{diag}\big( Q (Q\circ \Pi) \big) \label{Lambda2local}
\end{gather}
\end{subequations}
where 
$\circ$ indicates the Hadamard product, $Q = W_0^* (\tilde{\Delta} \otimes B)V_0 $.
The matrix $Q$ has block structure $Q = [[Q_{ik}]] $
where each block $[Q_{ik}] = \tilde{\Delta}_{ik} [R_{ik}]$ and the block $[R_{ik}] = {W_{0}^i}^* B V_{0}^k  \in \mathbb{C}^{n \times n}$. 
The matrix $\Pi$ has block structure in the blocks $[\Pi_{ik}] \in \mathbb{C}^{n \times n}$, with entries
\begin{equation} \label{eq:pi}
    [\Pi_{ik}]_{pq} = \begin{cases}
          0 \quad & p = q \ \& \ i = k \\ \dfrac{1}{\lambda_{0,i}^p - \lambda_{0,k}^q} \quad & \text{otherwise} 
    \end{cases}
\end{equation}
Equation\,\eqref{Lambda2local} is rewritten as in Eq.\,\eqref{widelocal}.
\begin{widetext}
\begin{equation} \label{widelocal}
\Lambda_2 
= \text{diag} \left(- \begin{bmatrix} 
    \tilde{\Delta}_{11} [R_{11}]  & \cdots & \tilde{\Delta}_{1n} [R_{1n}] \\
    \vdots & \ddots & \vdots \\
    \tilde{\Delta}_{n1} [R_{n1}] & \cdots & \tilde{\Delta}_{nn} [R_{nn}]
    \end{bmatrix} \begin{bmatrix}
    \tilde{\Delta}_{11} [R_{11}] \circ [\Pi_{11}] & \cdots & \tilde{\Delta}_{1n} [R_{1n}]\circ [\Pi_{1n}] \\
    \vdots & \ddots & \vdots \\
    \tilde{\Delta}_{n1} [R_{n1}] \circ[\Pi_{n1}] & \cdots & \tilde{\Delta}_{nn} [R_{nn}]\circ[\Pi_{nn}]
    \end{bmatrix}  \right)
\end{equation}
\end{widetext} 
The real part of the blocks on the main diagonal of $\Lambda_2$
are
\begin{equation} \label{BBlocal}
    \bar{\Lambda}_{2,i} = \textnormal{diag}\left( \sum_{k = 1}^m \tilde{\Delta}_{ik}^2 [U_{ik}] \right), 
\end{equation}
where the bar notation indicates the real part, and
\begin{equation} \label{eq:U}
    [U_{ik}] = -\textnormal{real}\left(\left({W_{0}^i}^* B V_{0}^k\right) \left(({W_{0}^k}^* B V_{0}^i ) \circ [\Pi_{ki}] \right) \right).
\end{equation}
The entries $\tilde \Delta_{ik}$ reflect information about the mismatches (and the eigenvectors of the Laplacian); the blocks $[U_{ik}]$ reflect information about the matrices $F$, $H$, the eigencouplings $\zeta_i=\sigma \gamma_i$ and $\zeta_k=\sigma \gamma_k$, and the matrix $B$ that determines which of the parameters inside the local dynamics is the mismatched parameter.

From Eq.\ \eqref{BBlocal}, we are interested in the sign of
the entries on the main diagonal of the block $[U_{ik}]$, i.e., $[U_{ik}]_{ss}$. We note that each block $[U_{ik}]$ can be parametrized in the blocks $F - \sigma \gamma_i H$, $F - \sigma \gamma_k H$ and $B$; and with knowledge of $F$, $H$ and $B$, it can be parametrized in the 
pair ($\zeta_i$, $\zeta_k$). This approach can be applied to study variations of all the eigenvalues of the matrix $\tilde{M}_0$.  However, in what follows, we focus on the eigenvalue(s) {of $\tilde{M}_0$ with the largest real part (either a single eigenvalue or a pair of complex conjugate eigenvalues), which we call {critical}. We first consider a transition A $\rightarrow$ S and label $i=1$ the block to which the critical eigenvalue(s) belongs. We call the real part of the critical eigenvalue(s) $c_0$, i.e., $c_0 = \max_j \ \bar{\lambda}_{0,1}^j$, and $s = \textnormal{argmax}_j \ \bar{\lambda}_{0,1}^j$.
Then Eq.\,\eqref{BBlocal} needs to be evaluated for $i=1$ and for a fixed value of $\zeta_1=a$, where $a$ is the lower MSF bound for the synchronization in the case of the identical oscillators. We approximate,
\begin{equation} \label{exp}
c(\epsilon)=c_0+\epsilon c_1 + \epsilon^2 c_2.
 \end{equation}
where $c(\epsilon)$ is the real part of the critical eigenvalue of $\Tilde{M}$ and the expansion coefficients $c_1$ and $c_2$ are characterized by $F$, $H$, $B$ and the eigencouplings $\sigma \gamma_i$. Both $c_1$ and $c_2$ are relevant to describe how $c$ varies with $\epsilon$, however in what follows we focus on the sign of $c_2$, which determines the curvature of $c(\epsilon)$.   If $c_2>0$ ($c_2<0$), we expect $c(\epsilon)$ to increase (decrease) with $\epsilon$, which results in a decrease (increase) of the synchronizability with parameter heterogeneity. 
If we find that $[U_{ik}]_{ss} <0$ ($[U_{ik}]_{ss} >0$) for all $\zeta_k \geq a$, then it follows that $c_2 < 0$ ($c_2 > 0$.) 
We remark that our main result is that we can characterize the improvement (hindrance) of 
the synchronizability in terms of  
{a parametric function that maps the $[U_{ik}]_{ss}$  to the variable $\zeta_k=\sigma \gamma_k$, which we call the Curvature Contribution Function.}
 A general conclusion is that the larger (the smaller) the Curvature Contribution Function, the more one can expect synchronizability to improve (worsen) as parameter heterogeneity increases.

We now consider the case that the functions $\bF$ and $\bH$ provide stability of the synchronous solution for $\sigma$ in a bounded interval $\sigma_{\min} \leq \sigma \leq \sigma_{\max}$, resulting in two transitions: an A $\rightarrow$ S transition for $\sigma \geq \sigma_{min}$ and an S $\rightarrow$ A transition for $\sigma \geq \sigma_{max}$.
Then, the same approach outlined above can be used to determine the effects of mismatches on $\sigma_{\max}$ in correspondence of the  S $\rightarrow$ A transition. In this case, we set $i = n$ and $\zeta_n = b$, where $b$ is the upper MSF bound for synchronization in the case of identical oscillators. Then $s = \textnormal{argmax}_j \ \bar{\lambda}_{0,n}^j$. Equation \eqref{exp} still holds with $c_1 = \lambda_{1,n}^s$ and $c_2= \lambda_{2,n}^s$. If for all $a \leq \zeta_k \leq b$, $[U_{nk}]_{ss}$ is positive (negative), then $c_2 > 0$ ($c_2 < 0$), indicating that mismatches enhance (reduce) the synchronizability independent of the topology and of the particular values of the mismatches.

\color{black}

As an example, we choose the individual oscillators to be  Chua systems,
\begin{equation} \label{eq:oscillators}
\begin{aligned}
   \bF (\bx)= \begin{bmatrix} \beta(y - x - g(x)) \\
    x - \alpha y + z \\
    \kappa y\end{bmatrix}, \  & \begin{cases}
      \beta = 10 \\ \kappa = -17.85 \\
      \alpha = 1
    \end{cases} \\
    g(x)\!=\!bx\!+\!\frac{b\!-\!a}{2}(|x\!-\!1|\!-\!|x\!+\!1|), \ & \begin{cases}
      a\!=\!-1.44 \\ b\!=\!-0.72 \end{cases}
\end{aligned}
\end{equation}
where ${\bx} = [x \ y \ z]^\top$ and we set the coupling function $\bH(\bx) = x$. 
We study the effect of mismatches in the parameter $\kappa$ with the nominal value shown in Eq.\,\eqref{eq:oscillators}. Hence, the constant matrices $F, B$ and $H$ that define this mismatched problem are:
\begin{equation} \label{eq:FH2}
\begin{gathered}
    F = \begin{bmatrix} \beta(-1 - b) & \beta & 0 \\
    1 & -\alpha & 1 \\
    0 & \kappa & 0 \end{bmatrix}, \\
    H = \begin{bmatrix}
        1 & 0 & 0 \\
    0 & 0 & 0 \\
    0 & 0 & 0
    \end{bmatrix}, \
    B = \begin{bmatrix}
        0 & 0 & 0 \\
        0 & 0 & 0 \\
        0 & 1 & 0
        \end{bmatrix}.
\end{gathered}
\end{equation}

 We study the sign of $c_2$ from Eq.\,\eqref{exp}. Since in the case of identical oscillators, the MSF predicts that there is only one transition A $\rightarrow$ S for $\zeta_1 \geq 6.00$,  we set $i = 1$. In Fig.\,\ref{fig:charac} we plot $[U_{1k}]_{ss}$ as we vary the eigencoupling $\zeta_k \geq 6.00$, from which we see that for all choices of $\zeta_k$, $[U_{1k}]_{ss} < 0$; therefore, $c_2 < 0$ and the synchronizability increases with the magnitude of {$\epsilon$ independent of the particular choices of the mismatches ($\delta_i$) and of the network topology (Laplacian $L$.) 
} \begin{figure}
     \centering
     \includegraphics[width=0.8\linewidth]{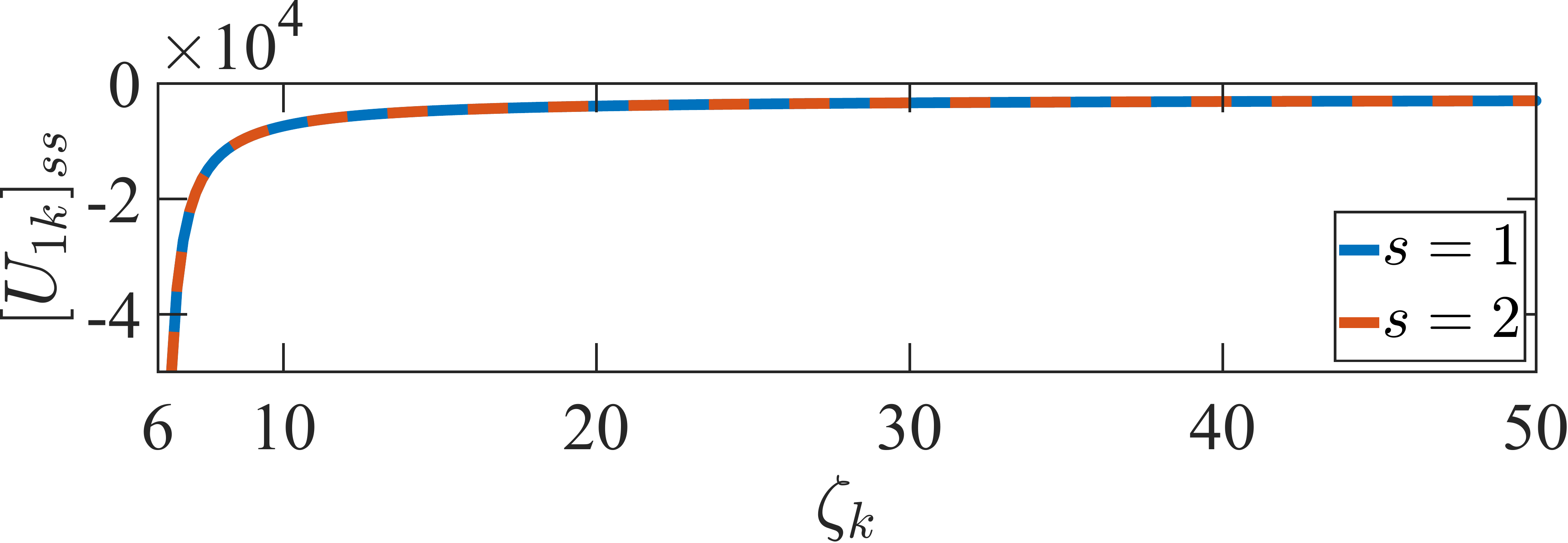}
     \caption{\textbf{{Curvature Contribution Function for Chua oscillators with parameter heterogeneity.}}
     The plot shows the real part of the first, and the second entry on the main diagonal of $[U_{1k}]$ versus $\zeta_k$. See Eq.\ \eqref{eq:FH2} for our choice of the matrices $F$, $H$, and $B$. 
    Since $[U_{1k}]_{ss} <0$ $\forall \zeta_k \geq 6.00$, then $c_2 <0$ in Eq.\,\eqref{exp}.
    }
     \label{fig:charac}
 \end{figure}
 
The result from Fig.\,\ref{fig:charac} is confirmed in Fig.\,\ref{fig:chua_params} where
 we plot the synchronization error (Fig.\,\ref{fig:chua_params}(a)) for a network of $N=3$ coupled Chua's oscillators (Fig.\,\ref{fig:chua_params}(b)) with randomly generated mismatches equal to $\delta_1 = -0.6534$, $\delta_2 = 0.7507$ and $\delta_3 = -0.0973$. The synchronization error is defined as
    $E = \left<\sqrt{ \frac{1}{N} \sum_{i=1}^N \left( x_i(t) - \bar{x}(t) \right)^2 }\right>,$
where $\bar x = 1/N \sum_{i=1}^N x_i$ is the average of the $x$ components of all oscillators, and $< \cdot >$ denotes the average over the time period $t$. 
Whenever nodes are approximately synchronized, $E$ attains a small value close to zero (the dark blue region.) The solid black curve delimits the region of stability predicted from the eigenvalues of $\tilde{M}$ from Eq.\,\eqref{eq:mtildelocal}. The dashed red curve is the approximation of the solid black curve obtained from second-order matrix perturbation theory.
\begin{figure}
    \centering
    \includegraphics[width=0.85\linewidth]{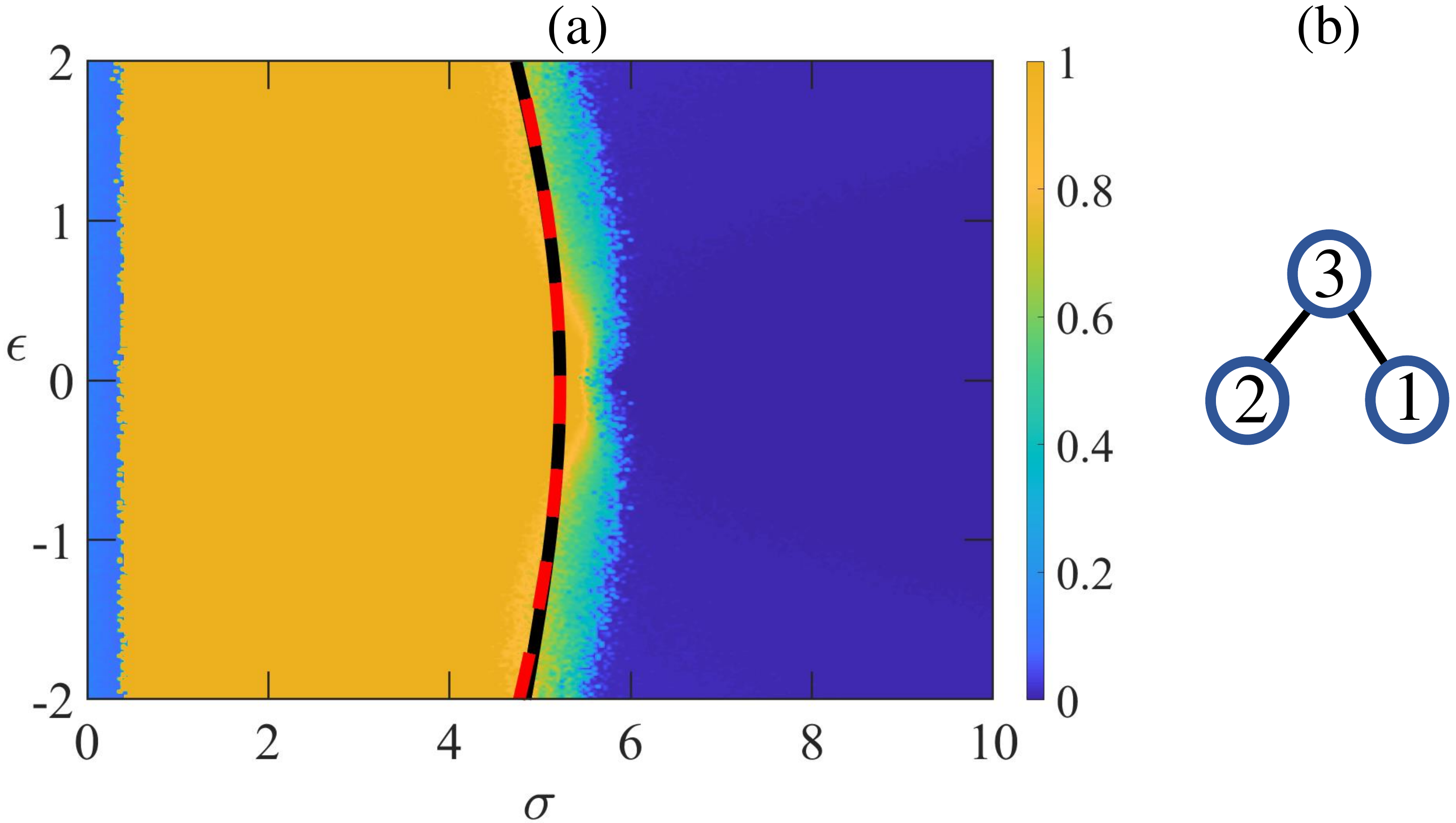}
    \caption{\textbf{Heterogeneity favors synchronization in networks of Chua oscillators.}
    (a) shows the synchronization error $E$ in the $(\sigma,\epsilon)$-plane for coupled Chua oscillators  with mismatches in the parameter $\kappa$, see Eq.\,\eqref{eq:oscillators}. The parameter $\sigma$ measures the coupling strength and the parameter $\epsilon$ measures the parameter heterogeneity. The solid black curve encloses the region of stability predicted from the eigenvalues of $\tilde{M}$ from Eq.\,\eqref{eq:mtildelocal}. The dashed red curve is the approximation of the solid black curve derived from matrix perturbation theory. (b) shows the topology of the network, with mismatches randomly set equal to
${\bdelta} =$  $[-0.6534, \allowbreak \
0.7507, \allowbreak \
-0.0973]^\top$.
    }
    \label{fig:chua_params}
\end{figure}
We note that the solid black contour from Fig.\,\ref{fig:chua_params} well predicts the transition to synchronization. 
Moreover, the solid black curve and its approximation given by the dashed red curve (the latter for small $|\epsilon|$) correctly replicate the curvature of the critical $\sigma$ versus the magnitude of the mismatches.

\subsection{Parametric Mismatches in the Frequencies of the Individual Oscillators} \label{sec:timescale}

In this section, we consider a problem studied in \cite{sugitani2021} in which mismatches affect the frequencies of the individual oscillators and show how our general theory presented above in the ``Parametric Mismatches in the Local Dynamics" subsection in the Results Section can be specialized to this particular problem.
We consider an equation for the time evolution of $N$ coupled dynamical systems, with frequency mismatches,
\begin{equation} \label{eq:timescale0}
      \dot{\bx}_i= (1+\epsilon {\delta}_i) \Big[   \bF(\bx_i)- \sigma  \sum_j {L}_{ij}\bH(\bx_j) \Big],
\end{equation}
$i=1,...,N,$ $\bx_i \in \mathbb{R}^m$ is the dynamical state of the node $i$,  $\bF : \mathbb{R}^m \mapsto \mathbb{R}^m$ and $\bH : \mathbb{R}^m \mapsto \mathbb{R}^m$ 
are
the local dynamics of each node and the coupling functions between the nodes, and $\sigma$ is a positive scalar measuring the strength of the coupling. Similar to the previous section, the symmetric Laplacian matrix $L=[L_{ij}]$ encodes the network connectivity.
The term $1+ \epsilon {\delta}_i$ is the particular frequency of  node $i$, with $1$ being the nominal frequency and $\epsilon {\delta}_i$ representing the possibly large frequency mismatch. As before, $\epsilon$ represents the tunable magnitude of the mismatches, and $\sum_i {\delta}_i = 0$, and $\sum_i {\delta}_i^2 = 1$. 
In Supplementary Note 3, we take arbitrary $\delta_i$'s and show how the equations can be rewritten in the form of Eq.\,\eqref{eq:timescale0}.

The average solution $\bar{\bx}(t)$ and the perturbations about the average solution $\bw_i (t)$ are defined as $\bar{\bx}(t) = 1/N \sum_i \bx_i(t)$ and $\bw_i (t) = \bx_i(t) - \bar{\bx}(t)$, $i = 1, \hdots, N$. Next, we rewrite Eq.\,\eqref{eq:timescale0} in $\bar{\bx}(t)$ and $\bw_i(t)$
by expanding to the first order about the average solution,
\begin{subequations} \label{eq:nonlinear1}
\begin{align}
\begin{split}
    \dot{\bar{\bx}} = \bF(\bar{\bx}) & + \frac{\epsilon}{N} \sum_j \Big[\delta_j D\bF(\bar{\bx}) \bw_j  - \sigma  d_j D\bH(\bar{\bx}) \bw_j\Big] 
\end{split}
\end{align}
\begin{align} 
\begin{split}
    \dot{\bw}_i  = & (1 + \epsilon\delta_i)D\bF(\bar{\bx}) \bw_i  - \frac{1}{N} \sum_j \epsilon\delta_j D\bF(\bar{\bx}) \bw_j  \\
    & -\sigma \sum_j (L_{ij} + \epsilon \delta_i L_{ij} + \frac{1}{N}\epsilon d_j) D\bH(\bar{\bx}) \bw_j   + \epsilon \delta_i  \bF(\bar{\bx}) 
\end{split}
\end{align}
\end{subequations}
where $D\bF, D\bH \in \mathbb{R}^{m \times m}$ are the time-varying Jacobian matrices of $\bF$ and $\bH$, and $d_j = \sum_i \delta_i L_{ij}$.
Local stability about the average solution is described the first order expansion  \eqref{eq:nonlinear1} under the assumption that the deviations ${\bw}_i$ are small.
By defining $\bW = [\bw_1 ^\top, \hdots, \bw_N ^\top]^\top \in \mathbb{R}^{Nm}$, we rewrite Eq.\,\eqref{eq:nonlinear1}, 
\begin{subequations} \label{eq:nonlinear2}
\begin{align}
\begin{split}
    \dot{\bar{\bx}} = \bF(\bar{\bx})   + & \frac{1}{N} \Big[\epsilon\bdelta^\top \otimes D\bF(\bar{\bx})  - \sigma  \epsilon \bd^\top  \otimes D\bH(\bar{\bx}) \Big] \bW  
\end{split}
\end{align}
\begin{align} \label{eq:W}
\begin{split}
    \dot{\bW}  = & \Big[(I_N + \epsilon\Delta_1 + \epsilon \Delta_2) \otimes D\bF(\bar{\bx})\Big] \bW   \\
    & -\sigma \Big[(L + \epsilon \Delta_1 L + \epsilon D) \otimes D\bH(\bar{\bx}) \Big] \bW   + \epsilon \Big(\bdelta \otimes \bF(\bar{\bx})\Big)
\end{split}
\end{align}
\end{subequations}
where $\bdelta = [\delta_1, \hdots, \delta_N]^\top$ and $\bd = [d_1, \hdots, d_N]^\top$, and the matrices $\Delta_1 = \textnormal{diag}({\delta}_1, ..., {\delta}_N)$,
\begin{equation}
\begin{gathered}
    \Delta_2 = \frac{-1}{N}\begin{bmatrix}
    {\delta}_1 & {\delta}_2 & \cdots & {\delta}_N\\
    {\delta}_1 & {\delta}_2 & \cdots & {\delta}_N\\
    \vdots & \vdots & \ddots & \vdots\\
    {\delta}_1 & {\delta}_2  & \cdots & {\delta}_N
    \end{bmatrix}, \,
    {D} = \frac{-1}{N}\begin{bmatrix}
    {d}_1 & {d}_2 & \cdots & {d}_N\\
    {d}_1 & {d}_2 & \cdots & {d}_N\\
    \vdots & \vdots & \ddots & \vdots\\
    {d}_1 & {d}_2  & \cdots & {d}_N
    \end{bmatrix}.
\end{gathered}
\end{equation}

As already noted before, Eq.\ \eqref{eq:W} is non-homogeneous, with the homogeneous part determining stability about the average solution and the non-homogeneous part determining the deviations of the individual trajectories from this solution, see also \cite{restr_bubbl,Su:Bo:Ni,SOPO,Amritkar2012,Amritkar2015}. We proceed similarly to what already done in the “Parametric Mismatches in the Local Dynamics” subsection in the Results and write the homogeneous part of Eq.\,\eqref{eq:W} in the form $\dot{\bW}  = M \bW$,
where $M = M_0 + \epsilon M_1$ and 
\begin{subequations} \label{eq:A}
\begin{gather} 
    M_0 = I_N \otimes F -\sigma L \otimes H \\
    M_1 = (\Delta_1 + \Delta_2) \otimes F - \sigma (\Delta_1 L + D) \otimes H
\end{gather}
\end{subequations}

  We multiply Eq.\,\eqref{eq:A} on the left by $(\tilde{V}^\top \otimes I)$ and on the right by $(\tilde{V} \otimes I)$ and obtain $\tilde{M}=\tilde{M}_0+\epsilon \tilde{M}_1$, where
\begin{subequations} \label{eq:mtilde}
\begin{equation}
    \tilde{M}_0 = (\tilde{V}^\top \otimes \tilde{V}) M_0 (\tilde{V} \otimes I)= I_{n} \otimes F - \sigma \tilde{\Gamma} \otimes H,
\end{equation}
\begin{align}
\begin{split}
    \tilde{M}_1 & = (\tilde{V}^\top \otimes I) M_1 (\tilde{V} \otimes I) \\
    & = \tilde{\Delta} \otimes F - \sigma \tilde{\Delta} \tilde{\Gamma} \otimes H  =(\tilde{\Delta} \otimes I ) \tilde{M}_0,
\end{split}
\end{align}
\end{subequations}
and we have used the property that $V^\top \Delta_2 V=V^\top {D} V= 0$, with $\tilde{M}, \tilde{M}_0, \tilde{M}_1 \in \mathbb{R}^{nm \times nm}$.

The rest of the analysis is done similarly to the ``Parametric Mismatches in the Local Dynamics” subsection in the Results. 
Next, we provide our final result analogous to Eq.\,\eqref{eq:U} for the case of mismatches in the frequencies.
Similar to Eq.\,\eqref{BBlocal}, the real part of the blocks on the main diagonal of $\Lambda_2$ are obtained,
\begin{equation} \label{BB}
    \bar{\Lambda}_{2,i} 
    = \textnormal{diag}\left( \sum_{k = 1}^n \tilde{\Delta}_{ik}^2 [U_{ik}] \right), 
\end{equation}
where each block 
\begin{equation}
    [U_{ik}] = -\textnormal{real}\left(\left({W_{0}^i}^* V_{0}^k \Lambda_{0}^k \right) \left(({W_{0}^k}^* V_{0}^i \Lambda_{0}^i) \circ [\Pi_{ki}] \right) \right).
\end{equation}
Here, $W_0^i$, $V_0^i$ and $\Lambda_0^i$ denote the left and right eigenvector matrices and a diagonal matrix with the eigenvalues of the $i$th block of $\tilde{M}_0$, respectively.
The block $[\Pi_{ki}]$ is defined the same as in Eq.\,\eqref{eq:pi}.
The entries $\tilde \Delta_{ik}$ reflect information about the mismatches (and the eigenvectors of the Laplacian), while the blocks $[U_{ik}]$ reflect information about the functions $\bF$, $\bH$, and the eigencouplings $\zeta_i=\sigma \gamma_i$ and $\zeta_k=\sigma \gamma_k$.

From Eq.\ \eqref{BB}, we are interested in the sign of the entries on the main diagonal of the block $[U_{ik}]$,i.e., $[{U}_{ik}]_{ss}$.
We note that each block $[U_{ik}]$ can be parametrized in the blocks $F - \sigma \gamma_i H$ and $F - \sigma \gamma_k H$; and with knowledge of $F$ and $H$, it can be parametrized in the pair ($\zeta_i$, $\zeta_k$). Next, we focus on the eigenvalue(s) {of $\tilde{M}_0$ with the largest real part (either a single eigenvalue or a pair of complex conjugate eigenvalues), which we call {critical}. We label $i=1$ the block to which the critical eigenvalue(s) belongs and expand the real part of the critical eigenvalue(s) to second order in 
For not too large $\epsilon$, $c(\epsilon)$ determines the crossing value of $\sigma_{\min}$ from asynchrony to synchrony. In particular, we focus on characterizing the curvature coefficient $c_2$.}

With the choice of $\bF$ and $\bH$ and having set $\zeta_1 = a$, we sweep over $\zeta_k$ and evaluate $[U_{1k}]_{ss}$.
If for all $\zeta_k \geq a$, $[U_{1k}]_{ss}$ is positive (negative) then 
$c_2 > 0$ ($c_2 < 0$), so the presence of mismatches reduces (increases) the synchronizability for any topology and any value of the mismatches.

Next, we consider two examples, one for which the range of stability increases with $\epsilon$ and one for which it decreases. The first example is that of Chua systems coupled in the `$x$' variable, considered in \cite{sugitani2021},
\begin{equation} \label{eq:chua}
\begin{gathered}
    \bF (\bx)= \begin{bmatrix} \eta(y - x - g(x)) \\
    x - y + z \\
    -y/k\end{bmatrix}, \   \begin{cases}
      \eta = 10 \\ k = 0.056
    \end{cases} \\
    g(x)\!=\!bx\!+\!\frac{1}{2}(b\!-\!a)(|x\!-\!1|\!-\!|x\!+\!1|), \  \begin{cases}
      a\!=\!-1.44 \\ b\!=\!-0.72. \end{cases}
\end{gathered}
\end{equation}
The matrices in Eq.\,\eqref{eq:A} are,
\begin{equation} \label{eq:FH}
\begin{gathered}
    F = \begin{bmatrix} \eta(-1 - b) & \eta & 0 \\
    1 & -1 & 1 \\
    0 & -1/k & 0 \end{bmatrix}, \quad 
    H = \begin{bmatrix} 1 & 0 & 0 \\
    0 & 0 & 0 \\
    0 & 0 & 0 \end{bmatrix}
\end{gathered}
\end{equation}
with the same parameters as in Eq.\,\eqref{eq:chua}.
With no parameter mismatches, the MSF \cite{Pe:Ca,Pecora2009} predicts the stability of the synchronous solution for $\zeta_1=\sigma \gamma_1 \geq a = 6$. We thus fix $\zeta_1 = a = 6$ and study the contribution of $\zeta_k \geq \zeta_1$ on $[U_{1k}]$. 
Figure \ref{fig:qhat} shows 
$[U_{1k}]_{ss}$ as a function of $\zeta_k \geq \zeta_1=6$. For any real positive $\zeta_k$, $F - \zeta_k H$ has one real and two complex and conjugate eigenvalues. We order these eigenvalues such that the first and the second eigenvalues are complex and conjugate and the third one is real. The complex eigenvalues are critical as these have real part larger than the real eigenvalue, so $s = 1,2$.  We characterize the second order coefficient of the expansion $c_2$ relative to these critical eigenvalues. Figure \ref{fig:qhat} shows that for all $\zeta_k  \geq \zeta_1=6$, 
$[U_{1k}]_{ss}$ 
is negative.  
Thus, based on Eq.\ \eqref{BB} we conclude that $c_2 < 0$.
\begin{figure}[ht]
    \centering\includegraphics[width=0.8\linewidth]{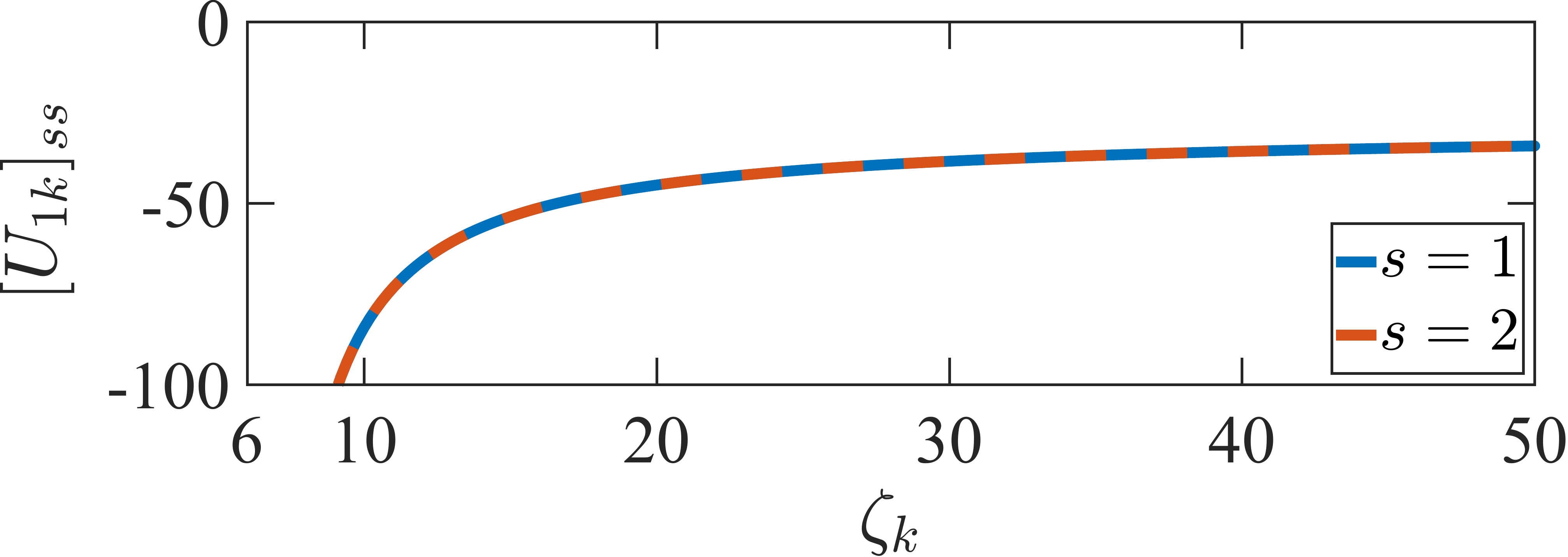}
    \caption{\textbf{{Curvature Contribution Function for Chua oscillators with heterogeneous frequencies.}} 
    We plot the Curvature Contribution Function,
    $[U_{1k}]_{ss}$, vs $\zeta_k$, $s=1,2$. In particular, the plot shows the real part of the first, and the second entry on the main diagonal of $[U_{1k}]$, respectively. See Eq.\,\eqref{eq:FH} for the matrices $F$ and $H$. For the case of coupled identical oscillators, synchronization is achieved for $\zeta_k \geq 6.00$. 
    }
    \label{fig:qhat}
\end{figure}
From Fig.\ \ref{fig:qhat} we see that
$c_2 < 0$ holds for any network topology, and so the enhancement of synchronization with the mismatches is a network-independent effect. 
Based on the results from Fig.\,\ref{fig:qhat}, we expect to see that increasing mismatches will improve the synchronizability for the case of Chua oscillators presented above.
This is confirmed in Fig.\,\ref{fig:chualinear} (a) where
we plot the synchronization error $E$ as the coupling strength $\sigma$ and the magnitude of the mismatches $\epsilon$ are varied for coupled Chua circuits with the undirected {network shown in Fig.\,\ref{fig:chualinear}(b). }
Whenever nodes are approximately synchronized, $E$ attains a small value close to zero (the blue region.) The black contour delimits the region of stability predicted from the eigenvalues of $\tilde{M}$.
\begin{figure}[ht]
    \centering
     {\includegraphics[width=0.75\linewidth]{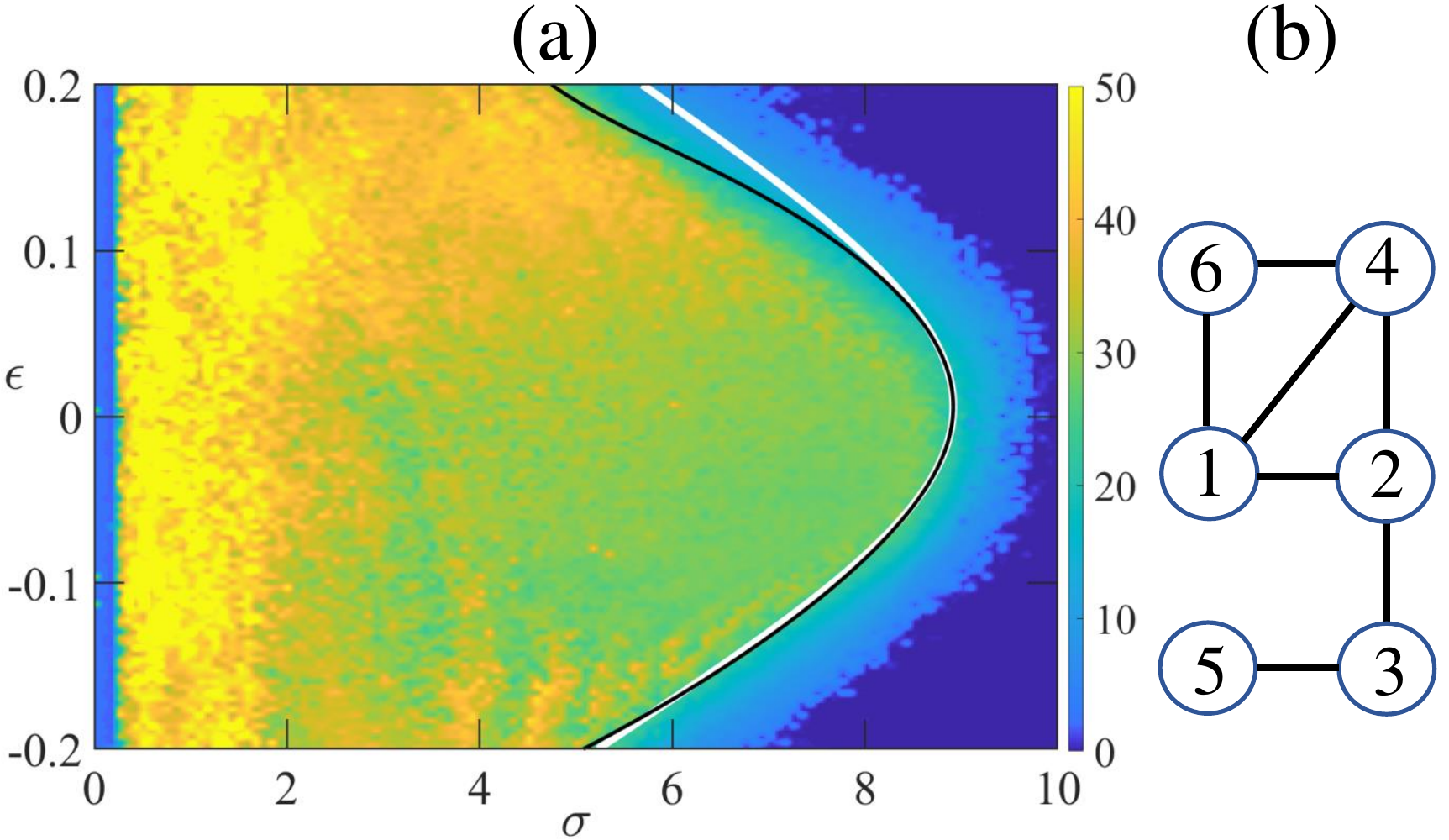}}
    \caption{\textbf{Analytical approximation of the region of stable synchronization 
    for Chua oscillators with heterogeneous frequencies.} (a) Synchronization error $E$ for a network of coupled Chua circuits, Eqs.\,\eqref{eq:timescale0} and \eqref{eq:chua}, as a function of the coupling strength $\sigma$ and the magnitude of the mismatches $\epsilon$. The black contour encloses the region of stability predicted from the eigenvalues of $\tilde{M}$ from Eq.\,\eqref{eq:mtilde}.   
    The white contour is an approximation of the black contour based on matrix perturbation theory, Eq.\,\eqref{exp}. (b) Topology of the network, with mismatches randomly set equal to ${\bdelta} =$  $[-0.067, \allowbreak \
-0.518, \allowbreak \
-0.358, \allowbreak \
0.712,  \allowbreak \
0.294, \allowbreak \
-0.062]^\top$. 
}  \label{fig:chualinear}
\end{figure}
We see that the linear approximation well describes the emergence of synchronization in the nonlinear system, Eq.\,\eqref{eq:timescale0}, even for relatively large mismatches, $|\epsilon| \simeq 0.2$. The 
linear approximation correctly predicts that the stability range increases for increasing $|\epsilon|$, as previously reported in \cite{sugitani2021}. 
The white curve in Fig.\,\ref{fig:chualinear}(a) is obtained by 
plotting $c(\epsilon)$ from Eq.\,\eqref{exp},
which provides a good approximation of the black curve, even for relatively large values of $|\epsilon|$.



The second example is that of Bernoulli maps with heterogeneous frequencies, described by discrete-time dynamics,
\begin{equation} \label{eq:discrete}
      {x}_i^{k+1}= \Bigg[ (1+\epsilon {\delta}_i) \Big[   f(x_i^k)- \sigma  \sum_j {L}_{ij} h(x_j^k) \Big] \Bigg] \textnormal{mod} \ 1, \ \ 
\end{equation}
$i=1,...,N,$ where $f(x^k) = 2 x^k$, and $h(x^k) = x^k$.
Figure\,\ref{fig:maps_exmp}(a) is a plot of the synchronization error versus $\epsilon$ and $\sigma$ {for a system of coupled maps with the undirected network in Fig.\,\ref{fig:maps_exmp}(b).
From Fig.\,\ref{fig:maps_exmp}(a),} we see that as $|\epsilon|$ grows the range of $\sigma$ that supports synchronization (blue area) shrinks. 
 For this case, $F=2$ and $H=1$. In Supplementary Note 4, we apply the same analysis presented before to approximate the eigenvalues of the matrix $\tilde{M}$ and obtain a more subtle result that most of the time $c_2 > 0$, but not always.
The region inside the black contour is where all the absolute values of the matrix $\tilde{M}$ are between $-1$ and $1$, which corresponds to the stability of the synchronous solution. The white contour is the approximation of the black curve based on the matrix perturbation theory. We see that the lower (upper) bound for $\sigma$ increases (decreases) with $|\epsilon|$, causing an overall reduction of the synchronizability with mismatches, which we find to be independent of the particular choice of the network topology (panel b).
\begin{figure}[ht]
    \centering
    \includegraphics[width = 0.9\linewidth]{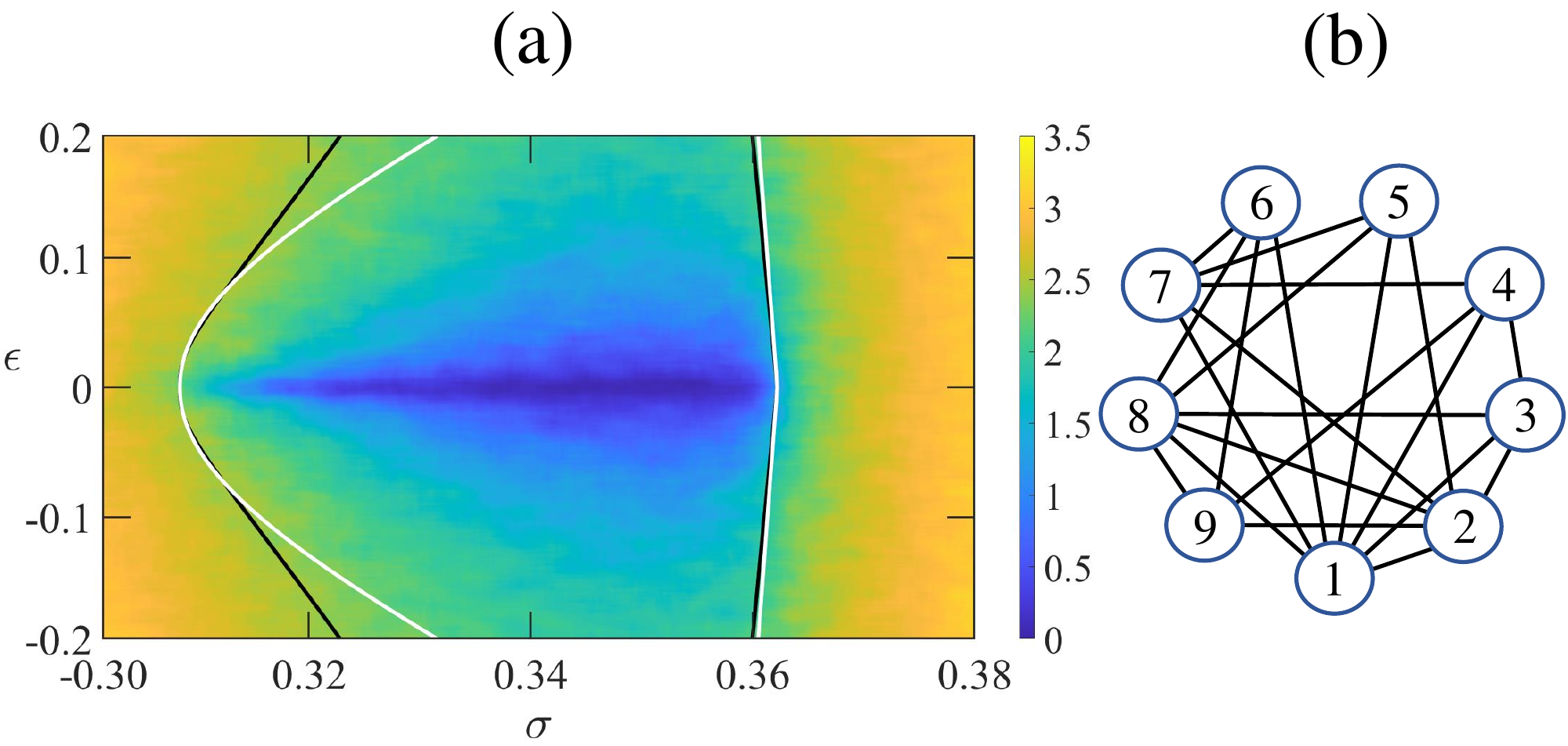}
    \caption{\textbf{Heterogeneity hinders synchronization in networks of Bernoulli maps.} (a) Synchronization error for a network of coupled Bernoulli maps, Eq.\,\eqref{eq:discrete}, as a function of the coupling strength $\sigma$ and the magnitude of the mismatches $\epsilon$. The region inside the black contours (eigenvalues) is where all the absolute values of the eigenvalues of $\tilde{M}$, Eq.\,\eqref{eq:mtilde}, are between -1 and 1. The white contour is an approximation of the black contour based on matrix perturbation theory. (b) Topology of the network. Here, the mismatches ${\bdelta} = [0.1568, \allowbreak
   -0.0869, \allowbreak
   -0.6469, \allowbreak
   -0.4689, \allowbreak
    0.0152, \allowbreak
    0.1642, \allowbreak
    0.3971, \allowbreak
    0.1033, \allowbreak
    0.3661]^\top$ were chosen randomly. \label{fig:maps_exmp}}
\end{figure}

We note that our analysis presented here for Chua oscillators and Bernoulli maps can be similarly applied
to different classes of oscillators and couplings. The case of R{\"o}ssler systems coupled in the $x$ or $y$ variables \cite{tang2022multilayer} is studied in Supplementary Note 5. The cases where $D\bF = D\bH$, including 
opto-electronic maps are presented in the next section.

\subsection{Discrete Time Opto-Electronic Systems}\label{sec:ex2}
We now consider a case for which the dynamics is in the general form of Eq.\,\eqref{eq:timescale0} and the functions $\bF$ and $\bH$ are taken in the particular form of discrete time opto-electronic systems \cite{hagerstrom2012experimental,NC, williams2013experimental},
\begin{align} \label{oe}
\begin{split}
     x_i^{k+1} = \Bigg[ (1 + \epsilon \delta_i) & \Big( \beta F(x_i^k) \\
      & - \sigma \sum_j L_{ij} F(x_j^k) + \alpha \Big) \Bigg] \ \text{mod} \ 2\pi,
\end{split}
\end{align}
where $F(x) = ({1 - \cos(x)})/{2}$, $\beta$ is the self-feedback strength, and the offset $\alpha$ is
introduced to suppress the trivial solution $x_i = 0$. The case of opto-electronic systems is of particular interest because our approach can be applied without introducing the assumption that the Jacobians are weakly dependent on the synchronous solution.

We assume possibly large mismatches and small perturbations about the average solution, $\bar{x}$. Then, by following steps similar to the previous sections,
we obtain the following equation that describes the time evolution of small transverse perturbations about the average solution,
\begin{equation} \label{oe2}
    \tilde{\bY}^{k+1}  =  DF(\bar{x}^k) \Bigg[(I_{N-1} + \tilde{\Delta}) (  \beta I_{N-1}  - \sigma \tilde{\Gamma} ) \Bigg] \tilde{\bY}^k .
\end{equation}
where $\tilde{\bY}^k \in \mathbb{R}^{N-1}$, and the matrices $\tilde{\Gamma}$ and $\tilde{\Delta}$ are exactly the same as they were defined in the previous sections. We define $\tilde{M} = (I_{N-1} + \tilde{\Delta}) (  \beta I_{N-1}  - \sigma \tilde{\Gamma} )$. By diagonalizing $\tilde{M}$,  Eq.\,\eqref{oe2} can be rewritten,
\begin{equation} \label{eq:DFperturb2}
     \bZ^{k+1}  =   \Phi DF(\bar{x}^k) \bZ^k
\end{equation}
where 
$\Phi$ is a diagonal matrix that has the eigenvalues of $\tilde{M}$ on its main diagonal. Equation (\ref{eq:DFperturb2}) can be decoupled into $N$ independent equations,
\begin{equation}
    z_i^{k+1}  =   \phi_i DF(\bar{x}^k) z_i^k, \quad i=1,\hdots,N.
\end{equation}
By defining the free parameter $s$, we can then study the stability of the parametric equation
\begin{equation} \label{eq:s}
    z^{k+1}  =   s DF(\bar{x}^k) z^k,
\end{equation}
as a function of the variable $s=\phi_i$. 
{For practical purposes, following \cite{Su:Bo:Ni,SOPO,panahi2021group}, the average solution $\bar{x}^k$ can be approximated by the dynamics of a single uncoupled map.}
We have calculated the Maximum Lyapunov Exponent (MLE) $\Psi(s)$ of  Eq.\,\eqref{eq:s} as a function of the real variable $s$ for the optoelectric map \cite{NC}, from which we have found that for $\beta = 2\pi$ and $\alpha = 0.525$, $\Psi(s)<0$ is negative in the range $[s^-=-3.47,s^+=3.47]$.

In Fig. \ref{fig:MLE_s} we plot the Maximum Lyapunov Exponent (MLE) $\Psi(s)$ of Eq.\,\eqref{eq:s} as a function of the real variable $s$ for the optoelectric map \cite{NC}. 
We see that $\Psi(s)<0$ is negative in the range $[s^-,s^+]$, and stability is achieved for $\phi^- > s^-$ and $\phi^+ < s^+$, where $\phi^-=\min_i \phi_i$ and $\phi^+=\max_i \phi_i$. Mismatches affect both $\phi^-$ and $\phi^+$, and we say that they enhance synchronization when they either increase $\phi^-$ or decrease $\phi^+$. 
\begin{figure}
    \centering
    \includegraphics[width=0.8\linewidth]{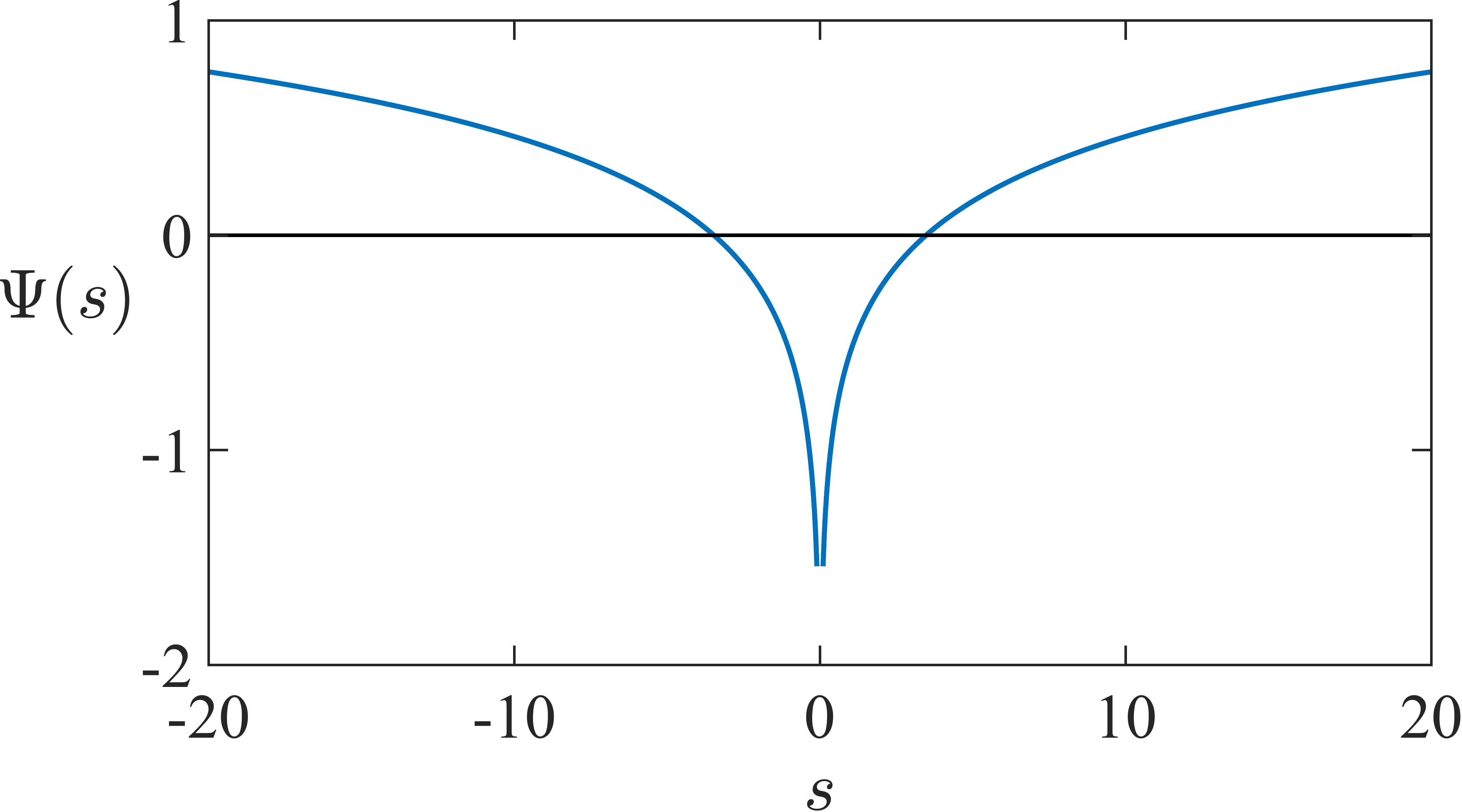}
    \caption{\textbf{Discrete Time Opto-Electronic Systems. Maximum Lyapunov Exponent for Stable Synchronization.} The  Maximum Lyapunov Exponent associated with Eq.\,\eqref{eq:s} as the parameter $s$ is varied. $\lim_{s \rightarrow 0} \Psi(s) = - \infty$. The zero-crossing points are $s^+= 3.47$, and $s^- = -3.47$. Here, the self-feedback
strength $\beta = 2\pi$ and the offset $\alpha = 0.525$.}
    \label{fig:MLE_s}
\end{figure}

\begin{figure*}
    \centering
    \includegraphics[width=0.95\textwidth]{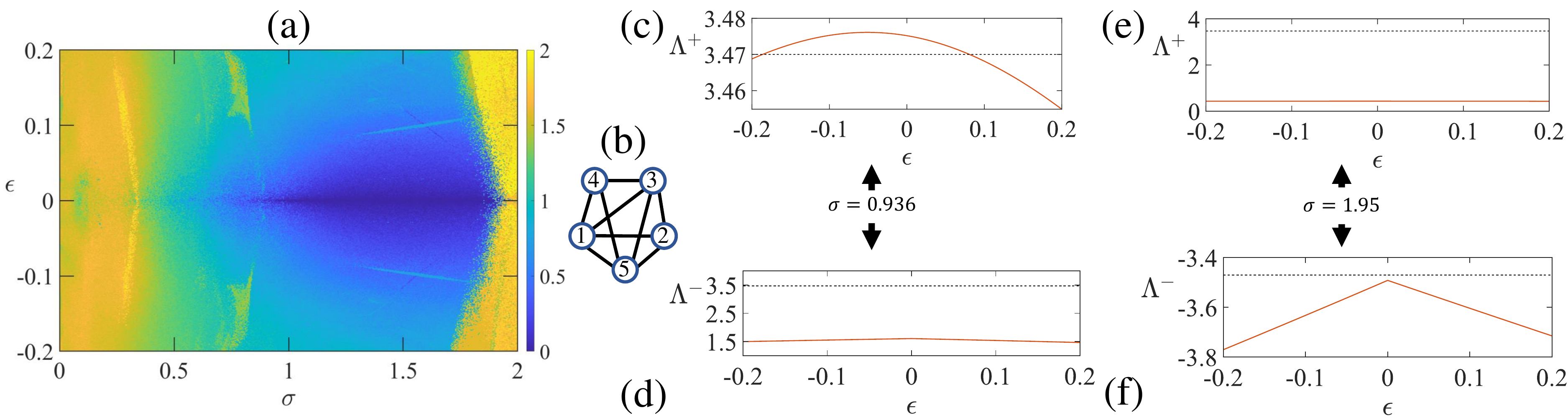}
    \caption{\textbf{The case of coupled `opto-electronic maps'.} (a) Synchronization error for a network of $N=5$ coupled optoelectric maps as a function of the magnitude of the mismatches $\epsilon$ and of the coupling strength $\sigma$. Here, the mismatches ${\bdelta} = [-0.5323, \allowbreak 
   -0.5265, \allowbreak 
    0.1526, \allowbreak 
    0.5058, \allowbreak 
    0.4003]^\top$ were randomly chosen. When the maps are identical ($\epsilon = 0$), synchronization is achieved for $\sigma^- = 0.94 \leq \sigma \leq 1.945 = \sigma^+$. We see that for $\sigma$ slightly lower than $\sigma^-$ ($\sigma$ slightly greater than 
    $\sigma^+$), the network of mismatched oscillators synchronizes (does not synchronize). (b) Topology of the network. (c) shows $\Lambda^+$, the maximum eigenvalue of $\tilde{M}$ as a function of $\epsilon$ for $\sigma = 0.936$. When $\epsilon = 0$, the network is not synchronized since $\Lambda^+$ is greater than the critical bound $s^+ = 3.47$ shown as a black line. As $\epsilon$ is either increased or decreased, $\Lambda^+$ eventually shifts below $s^+ = 3.47$ and the network synchronizes. (d) shows $\Lambda^-$, the minimum eigenvalue of $\tilde{M}$ remaining inside the critical bounds as $\epsilon$ varies for $\sigma = 0.936$. (e) shows $\Lambda^+$ remaining inside the critical bounds as  $\epsilon$ varies for $\sigma = 1.95$. (f) shows $\Lambda^-$ as a function of $\epsilon$ for $\sigma = 1.95$. When $\epsilon = 0$, $\Lambda^-$ is slightly less than the critical bound $s^-$. As $\epsilon$ is varied, $\Lambda^-$ shifts farther away from $s^-$; hence synchronization is not achieved.}
    \label{fig:opteo_error}
\end{figure*}

Next, we 
investigate how mismatches affect the eigenvalues of $\tilde{M}$. Once again, we use matrix perturbation theory to study variations of the eigenvalues of $\tilde{M}$ as the magnitude of the mismatches $\epsilon$ is varied. We write $\tilde{M} = \tilde{M}_0 + \epsilon \tilde{M}_1$, where
\begin{equation}
    \tilde{M}_0 = I_N - \sigma \Gamma, \quad \tilde{M}_1 = \tilde{\Delta} \tilde{M}_0,
\end{equation}
and $\Gamma$ is the diagonal matrix with diagonal entries equal to the eigenvalues of the Laplacian $L$. The matrix $\tilde{\Delta} = V^{-1} \Delta V$ where $V$ is a matrix whose columns are the eigenvectors of $L$. 
We have that the eigenvalues $\Lambda_\epsilon$ of the matrix $\tilde{M}$ can be approximated to second order, $ \Lambda_\epsilon =  \Lambda_0 + \epsilon \Lambda_1 + \epsilon^2 \Lambda_2$,
where the matrix $\Lambda_0$ is a diagonal matrix that has the eigenvalues of $A_0$ on its main diagonal (since $\tilde{M}_0$ is also diagonal, $\Lambda_0 = \tilde{M}_0$ and the matrices for the right and the left eigenvectors of $\tilde{M}_0$ are the identity) and,
\begin{subequations}
\begin{gather}
    {\Lambda_1=\mbox{diag}(\tilde{M}_1)=\mbox{diag}\left( \tilde{\Delta} \Lambda_0\right)}\\
    \Lambda_2 = - \textnormal{diag}\big( Q (Q\circ \Pi) \big). \label{Lambda2_maps}
\end{gather}
\end{subequations}
Here, $Q = [ Q_{ij}] = [ \tilde{\Delta}_{ij} \Lambda_{0,j} ]$ and
\begin{equation}
    \Pi_{ij} = \begin{cases}
          0 \quad & i = j \\ 
          \dfrac{1}{\Lambda_{0,i} - \Lambda_{0,j}} \quad & \text{otherwise} .
    \end{cases}
\end{equation}
Each $\Lambda_{2,i}$ is equal to,
\begin{equation} \label{eq:lambda2_maps}
    \Lambda_{2,i} = \sum_{j = 1}^N - \tilde{\Delta}_{ij} \Lambda_{0,j} (\tilde{\Delta}_{ji} \Lambda_{0,i} \circ \Pi_{ji}) = -\sum_{j=1}^N \tilde{\Delta}_{ij}^2 \tilde{\Pi}_{ji}.
\end{equation}
where
\begin{align}
\begin{split}
    \tilde{\Pi} & = \tilde{M}_0 \Pi \tilde{M}_0 = \begin{cases} \frac{(1-\sigma \gamma_i)(1-\sigma \gamma_j)}{\sigma \gamma_j - \sigma \gamma_i}  \quad & i \neq j \\
    0 & i = j \end{cases}.
\end{split}
\end{align}
We see that the sign of $\Lambda_{2,i}$ from Eq.\,\eqref{eq:lambda2_maps} is fully defined for a given pair of eigencouplings $\zeta_i = \sigma \gamma_i$ and $\zeta_j = \sigma \gamma_j$.
As discussed earlier, the network of coupled optoelectric maps with mismatches can synchronize if the eigenvalues of $\tilde{M}$ are in the range $[ -3.47, 3.47]$. Hence, the condition for synchronization is that,
\begin{equation}
   -3.47 \leq \Lambda_{0,i} + \epsilon \Lambda_{1,i} + \epsilon^2 \Lambda_{2,i} \leq 3.47 
\end{equation}
which is equivalent to
\begin{equation}
   -3.47 \leq (1 - \zeta_i) + \epsilon \tilde{\Delta}_{ii}(1 - \zeta_i) + \epsilon^2 \sum_{j=1}^N -\tilde{\Delta}_{ij}^2 \tilde{\Pi}_{ji} \leq 3.47 
\end{equation}

Figure \ref{fig:opteo_error} presents an example of application of our method. Panel (a) shows the synchronization error as a function of both $\sigma$ and $\epsilon$ for the $N=5$ network of coupled optoelectric maps shown in (b). For $\epsilon = 0$ we see that as $\sigma$ increases, there are two transitions: one around $\sigma = 0.94$, for which the identical systems switch from asynchrony to synchrony, and another one around $\sigma = 1.94$ where the transition is from synchrony to asynchrony. Panels (c-f) of Fig.\,\ref{fig:opteo_error}  show the application of our theory 
to predict whether or not mismatches enhance the network synchronizability. In this case, we see that both $\sigma^-$ and $\sigma^+$ decrease with $|\epsilon|$. In particular, panel (c) shows how $\Lambda^+$, the largest eigenvalue of $\tilde{M}$, varies with $\epsilon$ for $\sigma = 0.936$, which is slightly less than $\sigma^-=0.94$. We see that for either large enough $\epsilon$ or small enough $\epsilon$, $\Lambda^+$ decreases below $3.47$, which causes synchronization. Instead, panel (f) shows how $\Lambda^-$, the smallest eigenvalue of $\tilde{M}$, varies with $\epsilon$ for $\sigma = 1.95$, which is slightly more than $\sigma^+=1.945$. In this case, we see that as $|\epsilon|$ increases, $\Lambda^-$ decreases further away from $-3.47$, which is the threshold for synchronization.
{
Hence, we conclude that the presence of mismatches in the frequencies of the coupled opto-electronic maps enhances synchronization in the transition from asynchrony to synchrony (critical lower bound of $\sigma$) and hinders synchronization in the transition from synchrony to asynchrony (critical upper bound of $\sigma$.)
}

\section{Conclusions} \label{sec:conclusions}
This paper studies the emergence of synchronization in networks of coupled oscillators with possibly large parameter mismatches. We propose a parametric approach to analyze how the magnitude of the mismatches affects the synchronizability, i.e., the range of the coupling strength $\sigma$ for which the synchronous solution is stable. 
We identify a critical eigenvalue $c$ that is responsible for stability of the synchronous solution and expand $c$ to second order in the scalar $\epsilon$ which measures the magnitude of the mismatches; we focus on the second order coefficient of the expansion, i.e., the curvature coefficient $c_2$: if this coefficient is positive (negative), we expect the synchronizability to decrease (increase) with the size of the mismatches.   
{We derive an expression for how each pair of eigencouplings $(\zeta_i,\zeta_j)$ (the eigenvalues of the Laplacian matrix) contributes to the curvature coefficient $c_2$.

Similar to the parametric approach of the master stability function \cite{Pe:Ca}, it becomes possible, given knowledge of the functions $\bF$ and $\bH$, to pre-compute the effects of any pair of eigencouplings $(\zeta_i,\zeta_j)$,
seen as parameters,  {which is described by the Curvature Contribution Function}.} With this knowledge, one can then find for any network realization (any set of eigenvalues) the total curvature attained. For example, given knowledge of $\bF$ and $\bH$, our numerical approach consists of numerically computing the Curvature Contribution Function shown in Fig.\,\ref{fig:charac} and then from this, computing the overall effect on the curvature through Eq.\,\eqref{BBlocal}. 
In cases in which the Curvature Contribution Function is always positive (always negative), we can conclude that the curvature is positive (negative) independent of the network topology. Our main result is exactly that we provide an explanation for when parameter heterogeneity either increases or decreases the synchronizability.

Is the enhancement of synchronizability with parameter mismatches a general phenomenon as reported in \cite{sugitani2021}? We conclude that the answer is no, as our theory indicates that this is dependent on a number of factors and we find several examples of systems that become less synchronizable as we increase $|\epsilon|$. Our work applies to both continuous-time and discrete-time networks. 
{
The examples involving Chua oscillators showed that the presence of parametric mismatches improves synchronizability while the example of Bernoulli maps showed the opposite.
In the case of opto-electronic maps, it was shown that parameter heterogeneity had different effects on the synchronizability about different transitions: in the transition from asynchrony to synchrony (synchrony to asynchrony), the synchronizability was improved (hindered) by parameter heteorogeneity.
}

Our work differs substantially from Refs.\ \cite{restr_bubbl,Su:Bo:Ni,SOPO} since these papers are in the framework of small parameter mismatches $\leftrightarrow$ small state deviations. It also differs from  \cite{sugitani2021} as we
provide conditions, based on matrix perturbation theory, that allow us to explain why parameter mismatches sometimes favor synchronization and sometimes hinder it. We depart from the `mode mixing' hypothesis \cite{sugitani2021} and the implication that this should always result in an improvement in synchronizability. Such mode mixing (due to the presence of two non-commuting matrices \cite{boccaletti2006synchronization,HYP,Ir:So}) had already been observed in \cite{Amritkar2012,Amritkar2015} but no conclusions were drawn on its usefulness towards synchronization. Furthermore, our approach allows us to investigate the general case in which synchronization is possible in a bounded range of a given parameter and to consider the effects of parameter heterogeneity on both the lower and upper end of the range, which was not considered before.

\section*{Acknowledgements}
This work is supported by NIH grant 1R21EB028489-01A1. The authors thank Chad Nathe for insightful conversations and his help with performing simulations.

\section*{Data Availability}
The data that support the findings of this study are available
within the article.

\section*{Code Availability}
We will make the code available upon request.

\section*{Contributions}

A. N., S.P. and F.S. worked on the theory . A.N. worked on the numerical simulations. F.S. supervised the research.

\section*{Competing Interests}
The authors declare no competing interests.

\color{black}



 \newcommand{\noop}[1]{}

\end{document}